\documentclass[12pt]{article}
\title{Series representations for the Stieltjes constants} 
\author{Mark W. Coffey\\
Department of Physics\\
Colorado School of Mines\\
Golden, CO  80401\\
(Received $\mbox{~~~~~~~~~~~~~~~~~~~~~~~~~~~~~~~2009}$)}
\date{September 1, 2009}
\begin{document}
\maketitle
\baselineskip=25 pt
\begin{abstract}

The Stieltjes constants $\gamma_k(a)$ appear as the coefficients in the regular
part of the Laurent expansion of the Hurwitz zeta function $\zeta(s,a)$ about $s=1$.
We present series representations of these constants of interest to
theoretical and computational analytic number theory.  A particular result gives
an addition formula for the Stieltjes constants.  As a byproduct, expressions for
derivatives of all orders of the Stieltjes coefficients are given.  Many other
results are obtained, including instances of an exponentially fast converging
series representation for $\gamma_k=\gamma_k(1)$.  Some extensions are briefly described, as well as the relevance to expansions of Dirichlet $L$ functions.

\end{abstract}
 
\vspace{.25cm}
\baselineskip=15pt
\centerline{\bf Key words and phrases}
\medskip 

\noindent

Stieltjes constants, Riemann zeta function, Hurwitz zeta function, Laurent expansion, Stirling numbers of the first kind, Dirichlet $L$ functions, Lerch zeta function  

\vfill
\centerline{\bf 2000 AMS codes}
11M35, 11M06, 11Y60.  secondary:  05A10

\baselineskip=25pt
\pagebreak
\medskip
\centerline{\bf Introduction and statement of results}
\medskip

The Stieltjes (or generalized Euler) constants $\gamma_k(a)$ appear as 
expansion coefficients in the Laurent series for the Hurwitz zeta function 
$\zeta(s,a)$ about its simple pole at $s=1$ 
\cite{briggs,coffeyjmaa,hardy,kluyver,mitrovic}.   
These constants are important in analytic number theory and elsewhere, where
they appear in various estimations and as a result of asymptotic analyses.
They are also of much use in developing a binomial sum $S_\gamma(n)$
introduced by the author in the study of a critical subsum in application of
the Li criterion for the Riemann hypothesis \cite{coffey09}.
The constants $\gamma_k(1)$ are important in relation to the derivatives of 
the Riemann xi function $\xi(s)=\pi^{-s/2}\Gamma(s/2+1)(s-1)\zeta(s)$ at $s=1$, 
where $\Gamma$ is the Gamma function and $\zeta(s)$ is the Riemann zeta function
\cite{edwards,ivic,karatsuba,riemann}.  Hence some of the relevance to the Li
criterion.  

Despite the fact that many relations are known for the Stieltjes constants,
there are many open questions, including those concerned with their arithmetic
nature.  Even good estimations of their magnitudes is still lacking.  
In addition, formulas for obtaining the Stieltjes constants to arbitrary
precision remain of interest.

On the subject of the magnitudes $|\gamma_k(a)|$, known estimates \cite{berndt,yue}
are highly conservative, and there is much room for improvement.  We present one
approach for such estimation.  Our result is probably less important by itself,
than for the possibilities for extension and improvement that it suggests.

In this paper, we present various series representations of the Stieltjes
coefficients.  Most of these are fast converging, so have application to high
precision computation.  
In particular, we make use of the Stirling numbers of the first kind $s(k,j)$
and their properties.  Therefore, our treatment reflects a fusion of 
analytic number theory and enumerative combinatorics.  Moreover, one of our
Propositions (9) gives instances of an exponentially fast converging series
representation for $\gamma_k$.  Such very fast converging series are relatively 
rare, and this may be the first result of its kind for the Stieltjes constants.


The Hurwitz zeta function, initially defined by $\zeta(s,a)=\sum_{n=0}^\infty
(n+a)^{-s}$ for $\mbox{Re} ~s>1$, has an analytic continuation to the whole
complex plane \cite{berndt,titch,karatsuba}.   
In the case of $a=1$, $\zeta(s,a)$ reduces to the Riemann zeta function
$\zeta(s)$.  In this instance, by convention, the Stieltjes constants
$\gamma_k(1)$ are simply denoted $\gamma_k$ \cite{briggs,hardy,kluyver,kreminski,mitrovic,yue}.  We recall that $\gamma_0(a)=-
\psi(a)$, where $\psi=\Gamma'/\Gamma$ is the digamma function.  We also recall that
$\gamma_k(a+1)=\gamma_k(a)-(\ln^k a)/a$, and more generally that for $n \geq 1$ an
integer 
$$\gamma_k(a+n)=\gamma_k(a)-\sum_{j=0}^{n-1} {{\ln^k(a+j)} \over {a+j}}, \eqno(1.1)$$
as follows from the functional equation $\zeta(s,a+n)=\zeta(s,a)-\sum_{j=0}^{n-1}
(a+j)^{-s}$.

Unless specified otherwise below, letters $j$, $k$, $\ell$, $m$, and $n$ denote nonnegative integers.  We point out the common alternative notations $S_n^{(m)}$ and 
$\left[\begin{array}{c} n \\ m \end{array} \right]$
for Stirling numbers of the first kind $s(n,m)$ \cite{nbs,comtet,gkp}, with the relation 
$\left[\begin{array}{c} n \\ m \end{array} \right]=(-1)^{n+m}s(n,m)$ holding.  We also make use of $P_1(t) \equiv B_1(t-[t])=t-[t]-1/2$, the first periodized Bernoulli polynomial (e.g., \cite{ivic,titch}).  Obviously we have
$|P_1(t)| \leq 1/2$.

{\bf Proposition 1}.  (Addition formula for the Stieltjes constants) 
Let Re $a>0$ and $|a| > |b|$.  Then (i)
$$\gamma_\ell(a+b)=\gamma_\ell(a)+(-1)^\ell \sum_{j=2}^\infty {b^{j-1} \over {(j-1)!}}
\sum_{k=0}^\ell (-1)^k {\ell \choose k} s(j,k+1) k!\zeta^{(\ell-k)}(j,a).  \eqno(1.2)$$
(ii) We have
$$\gamma_\ell'(a)=(-1)^{\ell+1} [\zeta^{(\ell)}(2,a)+\ell \zeta^{(\ell-1)}(2,a)],
\eqno(1.3)$$
$$\gamma_\ell''(a)=(-1)^\ell[2\zeta^{(\ell)}(3,a)+3\ell \zeta^{(\ell-1)}(3,a) +\ell(\ell-1)\zeta^{(\ell-2)}(3,a)],  \eqno(1.4)$$  
and
$$\gamma_\ell'''(a)=(-1)^{\ell+1}[6\zeta^{(\ell)}(4,a)+11\ell \zeta^{(\ell-1)}(4,a) +6\ell(\ell-1) \zeta^{(\ell-2)}(4,a)+\ell(\ell-1)(\ell-2) \zeta^{(\ell-3)}(4,a)].  \eqno(1.5)$$  
\noindent
{\bf Corollary 1}.  
$$\gamma_1'(1)=\zeta(2)[\gamma+\ln(2\pi)+12\zeta'(-1)],  \eqno(1.6)$$
where $\zeta(2)=\pi^2/6$ and $\gamma=-\psi(1)$ is Euler's constant.

\noindent
{\bf Corollary 2}.  For $j \geq 1$ we have
$$\gamma_\ell^{(j)}(a)=(-1)^\ell \sum_{k=0}^\ell (-1)^k k!{\ell \choose k} s(j+1,k+1) \zeta^{(\ell-k)}(j+1,a).  \eqno(1.7)$$

{\bf Proposition 2}. 
(i) For Re $a>1$,
$$\gamma_n(a)=-{{\ln^{n+1}(a-1)} \over {n+1}}-n!\sum_{k=1}^\infty {1 \over {(k+1)!}} \sum_{m=0}^n {{(-1)^m} \over {m!}} s(k+1,n-m+1) \zeta^{(m)}(k+1,a), \eqno(1.8)$$
(ii) For Re $a>1/2$,
$$\gamma_n(a)=-{{\ln^{n+1}(a-1/2)} \over {n+1}}-n!\sum_{k=1}^\infty {1 \over {4^k(2k+1)!}}\sum_{m=0}^n {{(-1)^m} \over {m!}} s(2k+1,n-m+1) \zeta^{(m)}(2k+1,a), \eqno(1.9)$$
(iii) For Re $a>1/2$,
$$\gamma_n(a)=-{{\ln^{n+1}(a-1/2)} \over {n+1}}+n!\sum_{k=1}^\infty {{(-1)^k} \over {4^k(2k+1)!}}\sum_{m=0}^n {{(-1)^m} \over {m!}} s(2k+1,n-m+1) \zeta^{(m)}(2k+1,a)$$
$$-2n!\sum_{k=1}^\infty {1 \over {16^k(2k+1)!}}\sum_{m=0}^n {{(-1)^m} \over {m!}} s(4k+1,n-m+1) \zeta^{(m)}(4k+1,a).  \eqno(1.10)$$

{\bf Proposition 3}. (Asymptotic relation)
Let $B_j$ denote the Bernoulli numbers.  For Re $a>0$ and $\ell \to \infty$ we have
$$\gamma_\ell(a) \sim {1 \over {2a}}\ln^\ell a-{{\ln^{\ell+1}a} \over {\ell+1}} 
-\sum_{m=1}^\infty {B_{2m} \over {(2m)!}} a^{-2m} \sum_{k=0}^\ell 
{\ell \choose k} k! s(2m,k+1) \ln^{\ell-k}a.  \eqno(1.11)$$

{\bf Proposition 4}.  For any integer $N \geq 0$ and Re $a>0$ we have
$$\gamma_\ell(a)=\sum_{n=0}^N {{\ln^\ell (n+a)} \over {n+a}}-{{\ln^{\ell+1}(N+a)} 
\over {\ell+1}}+\sum_{r=2}^\infty {{(-1)^r} \over {r!}}\sum_{k=0}^\ell (-1)^k {\ell \choose k} k! s(r,k+1)\left[(-1)^\ell \zeta^{(\ell-k)}(r,a) \right.$$
$$\left. -(-1)^k\sum_{n=0}^N {{\ln^{\ell-k} (n+a)} \over {(n+a)^r}}\right].  \eqno(1.12)$$

{\bf Proposition 5}.  Let $a>0$.  Then there exists $a_1^*>1$ such that $\gamma_1(a)$
is monotonically increasing for $a<a_1^*$ and $\gamma_1(a)$ is monotonically
decreasing for $a>a_1^*$.  The approximate numerical value of $a_1^*$ is
$a_1^* \simeq 1.39112$.

{\bf Proposition 6}.  Let Re $z >0$.  Then we have
$${1 \over {j!}}\sum_{k=j}^\infty {z^k \over {(k-j)!}} \int_0^1 \gamma_k(a)da
= {{(-1)^j} \over z}.  \eqno(1.13)$$

{\bf Proposition 7}.  Put as in \cite{kreminski,yue}, $C_n(a) \equiv \gamma_n(a)-
{1 \over a}\ln^n a$ for $0 <a \leq 1$.  Then we have
$$|C_n(a)| \leq {{en!} \over {\sqrt{n} 2^n}}, ~~~~~~n \geq 1.  \eqno(1.14)$$


Our methods extend to many other analytic functions.  Additionally, although we 
present explicit results for expansions at $s=1$, this is not a restriction.
For those functions possessing functional equations, we also gain expansions
typically at $s=0$.  More generally, expansions about arbitrary points in the
complex plane are usually possible.

Another instance for which our methods apply is for the Lerch zeta function
$\Phi(z,s,a)$.   As an example, we consider the Lipshitz-Lerch transcendent
$$L(x,s,a)=\sum_{n=0}^\infty {e^{2\pi i nx} \over {(n+a)^s}}=\Phi(e^{2\pi i x},s,a)
,  \eqno(1.15)$$
for complex $a$ different from a negative integer.  We take in (1.15) $x$ real
and nonintegral, so that convergence obtains for Re $s>0$.  Else, for $x$ an
integer in (1.15), we reduce to the Hurwitz zeta function.  The functions $\Phi$
and $L$ possess integral representations and functional equations.  

A case of particular interest for (1.15) is when $x=1/2$.  Then we obtain the
alternating Hurwitz zeta function,
$$L\left({1 \over 2},s,a\right)=\sum_{n=0}^\infty {{(-1)^n} \over {(n+a)^s}}
=2^{-s}\left[\zeta\left(s,{a \over 2}\right)-\zeta\left(s,{{a+1} \over 2}\right)
\right].  \eqno(1.16)$$
Therefore, expansion at $s=1$ yields expressions for {\em differences} of Stieltjes
constants $\gamma_k(a/2)-\gamma_k[(a+1)/2]$.  More generally, for $x \notin Z$,
$L$ is nonsingular at $s=1$ and we may write
$$L(x,s,a)=\sum_{n=0}^\infty {{(-1)^n} \over {n!}} \ell_n(x,a)(s-1)^n.  \eqno(1.17)$$
We have
{\newline \bf Proposition 8}.  Let $a>0$ and $|\xi|<a$.  Then we have the addition
formula
$$\ell_n(x,a+\xi)=\ell_n(x,a)+\sum_{k=2}^\infty {\xi^{k-1} \over {(k-1)!}}\sum_{j=0}
^\infty (-1)^j {n \choose j}(n-j)!s(k,n-j+1) \left({d \over {ds}}\right)^j \left.
L(x,s+k-1,a)\right|_{s=1}.  \eqno(1.18)$$
As a result, we obtain for derivatives with respect to $a$ of $\ell_n$,
{\newline \bf Corollary 3}.
$$\ell_n^{(k)}(x,a)=\sum_{j=0}^\infty (-1)^j {n \choose j}(n-j)!s(k+1,n-j+1) \left(
{d \over {ds}}\right)^j \left. L(x,s+k,a)\right|_{s=1}.  \eqno(1.19)$$

Let $\Gamma(a,z)=\int_z^\infty t^{a-1}e^{-t}dt$ be the incomplete Gamma function,
$_pF_q$ the generalized hypergeometric function, Ei$(z)=-\int_{-z}^\infty e^{-t}
(dt/t)$ the exponential integral, and erf the error function \cite{nbs,grad}.
As a foretaste of a family of other results we offer
{\newline \bf Proposition 9}.  We have (i)
$${\gamma \over 2}=\sum_{n=1}^\infty {1 \over n} 
[1-\mbox{erf}(\sqrt{\pi}n)]-\sum_{n=1}^\infty \mbox{Ei}(-\pi n^2)-1+{1 \over 2} \ln(4\pi),  \eqno(1.20)$$
and (ii)
$$\gamma_1={\pi^2 \over {16}}+{\gamma \over 2}\psi \left({1 \over 2}\right)+{1 \over 8}
\psi^2\left({1 \over 2}\right)-1+{1 \over 2}\ln \pi-{1 \over 8}\ln^2 \pi$$
$$-{1 \over 2}\sum_{n=1}^\infty {1 \over {n}}\left[4n
 ~_2F_2\left({1 \over 2},{1 \over 2};{3 \over 2},{3 \over 2};-n^2 \pi\right)
+\psi\left({1 \over 2}\right)-2\ln n-\mbox{erf}(n\sqrt{\pi})\ln \pi  \right]$$
$$+\sum_{n=1}^\infty \left\{ {\gamma^2 \over 4}+{\pi^2 \over {24}}-{1 \over 2} n^2 \pi ~_3F_3(1,1,1;2,2,2;-n^2 \pi)+{1 \over 4}\ln(n^2 \pi)[2\gamma+\ln(\pi n^2)]
+{1 \over 2}\ln \pi \mbox{Ei}(-n^2\pi) \right \},  \eqno(1.21)$$
where $\psi(1/2)=-\gamma-2\ln 2$.

We introduce the polylogarithm function, initially defined by Li$_s(z)=\sum_{k=1}^
\infty {z^k \over k^s}$ for $|z| \leq 1$ and Re $s>1$, and analytically continued
thereafter.  We illustrate a method that more generally leads to expressions for
the sums $\gamma_k(a)+\gamma_k(1-a)$.  We have
{\newline \bf Proposition 10}.  Let $0<a<1$.  We have (i)
$$-\ln \pi+\psi\left({1 \over 2}\right)-\pi\cot \pi a-2\psi(a)=\gamma+\ln \pi+
2\left.{\partial \over {\partial s}}\right|_{s=0}[\mbox{Li}_s(e^{2\pi ia})+
\mbox{Li}_s(e^{-2\pi ia})], \eqno(1.22)$$
and (ii)
$$\gamma_1(a)+\gamma_1(1-a)={\pi^2 \over {12}}+{{\ln^2\pi} \over 4}-{{\ln \pi} \over
2}\psi\left({1 \over 2}\right)+{1 \over 4}\psi^2\left({1 \over 2}\right)+
{1 \over 2}\left[\ln \pi -\psi\left({1 \over 2}\right)\right][\psi(a)+\psi(1-a)]$$
$$-{1 \over 4}(\gamma+\ln \pi)^2-(\gamma+\ln \pi)\left.{\partial \over {\partial s}}\right|_{s=0}[\mbox{Li}_s(e^{2\pi ia})+\mbox{Li}_s(e^{-2\pi ia})]+
\left.{\partial^2 \over {\partial s^2}}\right|_{s=0}[\mbox{Li}_s(e^{2\pi ia})+
\mbox{Li}_s(e^{-2\pi ia})].  \eqno(1.23)$$

Lastly, we present expressions with Stirling numbers for rapidly converging
approximations to $\gamma=\gamma_0$.  Historically this subject \cite{ser} has 
been important especially due to P. Appell's use of them in his attempted proof of the 
irrationality of Euler's constant \cite{appell}.  In addition, we present an exact
representation for the difference between the Stieltjes constants and a finite sum.
Near the end of the paper, we further discuss these subjects.  We have
\newline{\bf Proposition 11}.  Define polynomials for $n \geq 1$
$$P_{n+1}(y)\equiv {1 \over {n!}}\int_0^y x(1-x)(2-x)\cdots (n-1-x) dx, \eqno(1.24)$$
and their values $p_{n+1} \equiv P_{n+1}(1)$.  Put $r_n^{(k)}=\gamma_k-D_n^{(k)}$,
where
$$D_n^{(k)} \equiv \sum_{m=1}^n {{\ln^k m} \over m}-{1 \over {k+1}}\ln^{k+1}(n+1).
\eqno(1.25)$$
Then we have (i)
$$P_{n+1}(y)={{(-1)^n} \over {n!}}\sum_{k=1}^n (-1)^k {{s(n-1,k-1)} \over {k(k+1)}}
[((k-1)y+y+1)(1-y)^{k-1}(y-1) + 1], \eqno(1.26)$$
(ii)
$$P_{n+1}(y)={{(-1)^{n+1}} \over {n!}}\sum_{k=0}^n {{s(n,k)} \over {k+1}}y^{k+1}$$
$$={{(-1)^{n+1}} \over {n!}}\left[\sum_{k=1}^{n-1} {{s(n,k)} \over {k+1}}y^{k+1}+
\delta_{n0} + {1 \over {n+1}}\right], \eqno(1.27)$$
where $\delta_{jk}$ is the Kronecker symbol, (iii) the special case
$$p_{n+1}={{(-1)^n} \over {n!}}\sum_{k=1}^n (-1)^k {{s(n-1,k-1)} \over {k(k+1)}}
={{(-1)^{n+1}} \over {n!}}\sum_{k=1}^n {{s(n,k)} \over {k+1}},  \eqno(1.28)$$
and (iv)
$$r_n^{(k)} = \sum_{m=n+1}^\infty I_{km}, \eqno(1.29a)$$
where
$$I_{km}=\sum_{n=1}^\infty {1 \over {(n+1)m^{n+1}}}\left[(-1)^n \ln^k m+{1 \over {n!}}
\sum_{j=0}^{k-1}{{k!} \over {(k-j-1)!}}s(n+1,j+2) \ln^{k-j-1}m\right].  \eqno(1.29b)$$

\bigskip
\centerline{\bf Proof of Propositions}
\medskip

{\it Proposition 1}.  We let $(z)_k=\Gamma(z+k)/\Gamma(z)$ be the Pochhammer symbol.
We make use of the following.
\newline{\bf Lemma 1}.  We have
$$\left.\left({d \over {ds}}\right)^\ell (s)_j \right|_{s=1}=(-1)^{j+\ell} \ell! s(j+1,\ell+1).  \eqno(2.1)$$
We have the standard expansion
$$(s)_j=s(s+1)\cdots (s+j-1)=\sum_{k=0}^j (-1)^{j+k} s(j,k) s^k, \eqno(2.2)$$
giving
$$\left({d \over {ds}}\right)^\ell (s)_j =\sum_{k=\ell}^j (-1)^{j+k} s(j,k) 
k(k-1)\cdots (k-\ell+1) s^k, \eqno(2.3)$$
so that
$$\left.\left({d \over {ds}}\right)^\ell (s)_j \right|_{s=1}=(-1)^j \sum_{k=\ell}^j
(-1)^k s(j,k) (-1)^{\ell+1} k (1-k)_{\ell-1}$$
$$=(-1)^j (\ell-1)!\sum_{k=\ell}^j (-1)^k s(j,k)k{{k-1} \choose {\ell-1}}$$
$$=(-1)^j \ell!\sum_{k=\ell}^j (-1)^k s(j,k){k \choose \ell}.  \eqno(2.4)$$
It is known that \cite{gkp} (p. 265)
$$\sum_{k=\ell}^j (-1)^k s(j,k){k \choose \ell}=(-1)^\ell s(j+1,\ell+1).  \eqno(2.5)$$
Therefore, Lemma 1 follows.  

In order to obtain Proposition 1 we apply a formula of Wilton \cite{wilton2} for the Hurwitz zeta function,
$$\zeta(s,a+b)=\zeta(s,a)+\sum_{j=1}^\infty {{(-1)^j} \over {j!}} {{\Gamma(s+j)}
\over {\Gamma(s)}}\zeta(s+j,a) b^j, ~~s \neq 1, ~~|b| <|a|, ~~\mbox{Re}~a >0,
\eqno(2.6)$$
along with the product rule
$$\left.\left({d \over {ds}}\right)^\ell (s)_j \zeta(s+j,a)\right|_{s=1}=\left.
\sum_{k=0}^\ell {\ell \choose k} \left[\left({d \over {ds}}\right)^k (s)_j\right]
\zeta^{(\ell-k)}(s+j,a)\right|_{s=1}.  \eqno(2.7)$$
The defining Laurent expansion for the Stieltjes constants is 
$$\zeta(s,a)={1 \over {s-1}}+\sum_{k=0}^\infty {{(-1)^k \gamma_k(a)} \over
k!} (s-1)^k, ~~~~~~ s \neq 1. \eqno(2.8)$$
Therefore, the Wilton formula (2.6) gives
$$\sum_{k=0}^\infty {{(-1)^k \over k!} \gamma_k(a+b)} (s-1)^k=\sum_{k=0}^\infty 
{{(-1)^k \over k!} \gamma_k(a)} (s-1)^k +\sum_{j=1}^\infty {{(-1)^j} \over {j!}} 
(s)_j \zeta(s+j,a)b^j, $$
$$~~~~~~~~~~~~~~~~~~~~~~~~~~~~~~~~~~~~~~~~ ~~|b| <|a|, ~~\mbox{Re}~a >0.  \eqno(2.9)$$
We take $\ell$ derivatives of this equation and use (2.7) and Lemma 1, yielding
$$(-1)^\ell\gamma_\ell(a+b)=(-1)^\ell \gamma_\ell(a)+\sum_{j=1}^\infty {b^j \over
{j!}}\sum_{k=0}^\ell {\ell \choose k} (-1)^k s(j+1,k+1)k! \zeta^{(\ell-k)}(j+1,a).
\eqno(2.10)$$
Therefore, Proposition (i) follows.

For part (ii), we form the limit difference quotient
$$\gamma_\ell'(a)=\lim_{b \to 0} {1 \over b}[\gamma_\ell(a+b)-\gamma_\ell(a)],
\eqno(2.11)$$
giving
$$\gamma_\ell'(a)=(-1)^\ell \sum_{k=0}^\ell (-1)^k {\ell \choose k}s(2,k+1)k!
\zeta^{(\ell-k)}(2,a).  \eqno(2.12)$$
The Stirling number truncates the summation, with $s(2,k+1)=(-1)^{k+1}$, $k=0,1$,
and otherwise is $0$ for $k \geq 2$.  

We next have
$$\gamma_\ell''(a)=\lim_{b \to 0} {1 \over b^2}[\gamma_\ell(a+2b)-2\gamma_\ell(a+b)
+\gamma_\ell(a)]$$
$$=\lim_{b \to 0} {1 \over b^2}\{[\gamma_\ell(a+2b)-\gamma_\ell(a)]-2
[\gamma_\ell(a+b)-\gamma_\ell(a)]\}$$
$$=(-1)^\ell[2\zeta^{(\ell)}(3,a)+3\ell \zeta^{(\ell-1)}(3,a)+\ell(\ell-1)
\zeta^{(\ell-2)}(3,a)].  \eqno(2.13)$$  
Similarly higher derivatives of $\gamma_\ell(a)$ may be determined and part (ii) 
has been shown.  

From part (ii), we have $\gamma_1'(1)=\zeta'(2)+\zeta(2)$, wherein by the
functional equation of the Riemann zeta function, or otherwise, we have the
relations
$$\zeta'(2)=\zeta(2)(\gamma+\ln 2-12\ln A+\ln \pi)$$
$$=\zeta(2)[\gamma+\ln(2\pi)-1+12 \zeta'(-1)], \eqno(2.14)$$
where $\ln A=1/12-\zeta'(-1)$ and $A$ is Glaisher's constant.  Corollary 1 follows.

For Corollary 2 we simply shift the summation index $j \to j+1$ in the Taylor series (1.2) and read off the derivatives.  Otherwise, we could make use of the forward
difference operator $b^{-n}\Delta_b^n [f](x)=b^{-n}\sum_{k=0}^n {n \choose k}
(-1)^{n-k} f(x+kb)$.

{\bf Remarks}.  The auxiliary relation (2.5) can be proved in a number of ways.
One is with induction by using a recursion relation satisfied by $s(j,k)$ and
by the binomial coefficient.  
Other methods include the use of integral
representations either for the Stirling numbers of the first kind or for the
binomial coefficient.  For a proof using boson operators, see \cite{katriel}
(Identity 1).  

It is readily checked that at $\ell=0$ Proposition 1 (i) yields the identity
$\gamma_0(a+b)=-\psi(a+b)$.  For we have
$$\gamma_0(a+b)=\gamma_0+\sum_{j=1}^\infty (-b)^j \zeta(j+1,a)$$
$$=\gamma_0(a)+\sum_{j=1}^\infty {{(-b)^j} \over {j!}}\int_0^\infty {{t^j e^{-(a-1)t}}
\over {e^t-1}} dt$$
$$=\gamma_0(a)+\int_0^\infty {{e^{-at} (e^{-bt}-1)} \over {1-e^{-t}}}dt$$
$$=\gamma_0(a)+\psi(a)-\psi(a+b)=-\psi(a+b). \eqno(2.15)$$
Herein we used a standard integral representation for $\zeta(s,a)$ and for the
digamma function \cite{nbs} (p. 259) or \cite{grad} (p. 943).

Similarly, we may write
$$\gamma_1(a+b)=\gamma_1(a)-\sum_{j=1}^\infty (-b)^j[\zeta'(j+1,a)+H_j \zeta(j+1,a)],
\eqno(2.16)$$
where $H_j \equiv \sum_{k=1}^j {1 \over k}$ is the usual harmonic number.
Again, integral forms of this relation may be given.


Our integral representation \cite{coffeystv2} [Proposition 3(a)],
$$\gamma_k(a)= {1 \over {2a}}\ln^k a-{{\ln^{k+1} a} \over {k+1}} + {2 \over a}
\mbox{Re} \int_0^\infty {{(y/a-i)\ln^k (a-iy)} \over {(1+y^2/a^2)(e^{2 \pi y}
-1)}}dy, ~~~~\mbox{Re} ~ a>0,  \eqno(2.17)$$   
could also be used to prove Proposition 1 part (ii).

The approximate numerical value $\gamma_1'(1) \simeq 0.707385812532$ suggests
that $\gamma_1'(1)=\zeta'(2)+\zeta(2)$ can be written as $1/\sqrt{2}$ together with
a series of systematic correction terms.  

Of course, the Wilton formula has built in the relation ${\partial \over {\partial a}}
\zeta(s,a)=-s\zeta(s+1,a)$, and more generally that
$$\left({\partial \over {\partial a}}\right)^j \zeta(s,a)=(-1)^j (s)_j\zeta(s+j,a).
\eqno(2.18)$$



{\it Proposition 2}.  
These results are based upon \cite{coffeyjcam08} (Proposition 1, parts (i), (ii), and
(iv)).  We have for each indicated domain of $a$,
$$\zeta(s,a)={{(a-1)^{1-s}} \over {s-1}} -{1 \over {\Gamma(s)}}\sum_{k=1}^\infty
{{\Gamma(s+k)} \over {(k+1)!}} \zeta(s+k,a), ~~~~\mbox{Re} ~a>1,  \eqno(2.19)$$
$$\zeta(s,a)={{(a-1/2)^{1-s}} \over {s-1}} -{1 \over {\Gamma(s)}}\sum_{k=1}^\infty
{{\Gamma(s+2k)} \over {4^k(2k+1)!}} \zeta(s+2k,a), ~~~~\mbox{Re} ~a>1/2, \eqno(2.20)$$
and
$$\zeta(s,a)=2^{s-2}{{(2a-1)^{1-s}} \over {s-1}} +{1 \over {\Gamma(s)}}
\left[ {1 \over 2} \sum_{k=0}^\infty {{(-1)^k\Gamma(s+2k)} \over {4^k(2k+1)!}}
 \zeta(s+2k,a)\right.$$
$$\left.-\sum_{k=1}^\infty {{\Gamma(s+4k)} \over {16^k(4k+1)!}} \zeta(s+4k,a)
\right], ~~~~\mbox{Re} ~a>1/2.   \eqno(2.21)$$
Lemma 1 immediately carries over to
$$\left.\left({d \over {ds}}\right)^\ell (s)_{pj} \right|_{s=1}=(-1)^{pj+\ell} \ell! s(pj+1,\ell+1), ~~~~p \geq 1.  \eqno(2.22)$$
We then expand equations (2.19)-(2.21) in powers of $s-1$ using the product rule and
the Proposition follows.  In the case of part (iii), we have first moved the $k=0$ 
term on the right side of Eq. (2.21) to the left side and multiplied the resulting 
equation by $2$.

{\bf Remarks}.  Part (ii) of Proposition 2 gives the general $n$ case beyond the
low order instances given explicitly in terms of generalized harmonic numbers 
$H_n^{(r)}$ in Proposition 7 of Ref. \cite{coffeystdiffs}.  After all, it is well
known that $s(n+1,1)=(-1)^n n!$, $s(n+1,2)=(-1)^{n+1}n! H_n$, and $s(n+1,3)
=(-1)^n n![H_n^2-H_n^{(2)}]/2$, where $H_n \equiv H_n^{(1)}$.  This part of the
Proposition has also been obtained by Smith \cite{rsmith}.  

Part (i) of Proposition 2 exhibits slow convergence.  In contrast, parts (ii) and
(iii) are very attractive for computation.  

The Ma\'{s}lanka representation for the Riemann zeta function is written in terms of
certain Pochhammer polynomials $P_k(s)$ \cite{maslanka01}.  Therefore, it is also possible to write expressions for $\gamma_j$ from this representation in terms of
sums including Stirling numbers of the first kind.  

Likewise, the Stark-Keiper formula for $\zeta$ may be used to develop expressions
for the Stieltjes constants.  For $N >0$ an integer, this representation reads
$$\zeta(s,N)=-{1 \over {s-1}}\sum_{k=1}^\infty \left(N+{{s-1} \over {k+1}}\right) {{(-1)^k} \over {k!}} (s)_k \zeta(s+k,N).  \eqno(2.23)$$  

Serviceable series for the Stieltjes constants using the Stirling numbers of
the first kind can also be written using the Taylor-series based expressions
$$\zeta(s,a)=a^{-s}+\sum_{n=0}^\infty {{(-a)^n} \over {n!}} (s)_n \zeta(s+n),
~~~~~~|a|<1, \eqno(2.24)$$  
and
$$\zeta\left(s,a+{1 \over 2}\right)=\sum_{n=0}^\infty {{(-a)^n} \over {n!}} (s)_n (2^{s+n}-1)\zeta(s+n),  ~~~~~~|a|<1/2.  \eqno(2.25)$$  

In general, Dirichlet $L$ functions may be written as a combination of Hurwitz 
zeta functions.  For instance, for $\chi$ a character modulo $m$ and Re $s \geq 1$
we have
$$L(s,\chi) = \sum_{k=1}^\infty {{\chi(k)} \over k^s} ={1 \over m^s}\sum_{k=1}^m 
\chi(k) \zeta\left(s,{k \over m}\right).  \eqno(2.26)$$
For $\chi$ a nonprincipal character, convergence obtains herein for Re $s \geq 0$.
Therefore, our results are very pertinent to derivatives and expansions of
Dirichlet $L$ series about $s=1$.  Especially for real-Dirichlet-character 
combinations of low order Stieltjes constants, Ref. \cite{coffeystdiffs} may be
consulted.

{\it Proposition 3}.  We have the representation valid for Re $s>-(2n-1)$,
$$\zeta(s,a)=a^{-s} + {a^{1-s} \over {s-1}}  
+\sum_{k=1}^n (s)_{k-1} {B_k \over {k!}}a^{-k-s+1} ~~~~~~~~~~~~~~~~~~~~~~~~~~~~~~~$$
$$~~~~~~~~~~~~~~~~~~~~ + {1 \over {\Gamma(s)}}\int_0^\infty \left ({1 \over {e^t-1}}
-\sum_{k=0}^n {B_k \over {k!}} t^{k-1}\right) e^{-at} t^{s-1}dt.  \eqno(2.27)$$
If we take $n \to \infty$ in this equation we in fact obtain an analytic
continuation of the Hurwitz zeta function to the whole complex plane.
We then develop the result,
$$\zeta(s,a)={1 \over 2}a^{-s} + {a^{1-s} \over {s-1}}  
+\sum_{j=2}^\infty (s)_{j-1} {B_j \over {j!}}a^{-j-s+1}, \eqno(2.28)$$
as a series in powers of $s-1$, where simply the factor
$$\left.\left({d \over {ds}}\right)^{\ell-k}a^{-j-(s-1)}\right|_{s=1}
=a^{-j}(-1)^{\ell-k} \ln^{\ell-k} a.  \eqno(2.29)$$
We use the product rule for derivatives of the summation term together with 
Lemma 1 and find 
$$\gamma_\ell(a) \sim {1 \over {2a}}\ln^\ell a-{{\ln^{\ell+1}a} \over {\ell+1}} 
+\sum_{j=2}^\infty (-1)^{j-1} {B_j \over {j!}} a^{-j} \sum_{k=0}^\ell 
{\ell \choose k} s(j,k+1)k! \ln^{\ell-k}a.  \eqno(2.30)$$
Since $B_{2n+1} = 0$ for $n \geq 1$, the stated form of the Proposition follows.

{\it Proposition 4}.  We employ the representation \cite{apostol} (p. 270)
based upon Euler-Maclaurin summation and integration by parts, for $N\geq 0$,
and Re $s >-m$, with $m=1,2,\ldots$,
$$\zeta(s,a)=\sum_{n=0}^N {1 \over {(n+a)^s}}+{{(N+a)^{1-s}} \over {s-1}}
-\sum_{r=1}^m {{(s)_r} \over {(r+1)!}} \left[\zeta(s+r,a)-\sum_{n=0}^N {1 \over
{(n+a)^{s+r}}}\right]$$
$$-{{(s)_{m+1}} \over {(m+1)!}}\sum_{n=N}^\infty \int_0^1 {u^{m+1} \over {(n+a+u)^
{s+m+1}}}du.  \eqno(2.31)$$
Taking $m \to \infty$ and developing the result in powers of $s-1$ using Lemma 1
gives the Proposition.

{\bf Remarks}.  In practice in using Proposition 4, there will be a tradeoff 
in selecting $N$ and the cutoff or otherwise estimating the remainder neglected
in the sum over $r$.  We expect the convergence to be poor as Re $a \to 0$ in this
Proposition.  We anticipate that a suitable procedure for computations is to
take values with Re $a>1$, and then to use relation (1.1) as needed.

Similarly, we could employ the representation \cite{grad} (p. 1073) for integers
$N \geq 0$ and Re $a>0$
$$\zeta(s,a)=\sum_{n=0}^N {1 \over {(n+a)^s}}+{{(N+a)^{1-s}} \over {s-1}}
-s\sum_{n=N}^\infty \int_n^{n+1} {{(t-n)} \over {(t+a)^{s+1}}}dt, ~~~\mbox{Re} ~s >1,
\eqno(2.32)$$
where the integral may be easily expressed in closed form.  Or we could use the
representation for integers $N \geq 0$ and Re $a>0$ \cite{karatsuba} (p. 16)
$$\zeta(s,a)=\sum_{n=0}^N {1 \over {(n+a)^s}}+{1 \over {s-1}}\left(N+{1 \over 2}+ a\right)^{1-s}+s\int_{N+1/2}^\infty {{P_1(t)} \over {(t+a)^{s+1}}}dt, ~~~\mbox{Re} 
~s >0.  $$
As we easily have that the integral in this equation is bounded by
$$|s|\left|\int_{N+1/2}^\infty {{dt} \over {(t+a)^{s+1}}}\right|={1 \over {|N+1/2+a|^s}},$$
the integral converges uniformly for $s$ in any compact subset of the half plane
Re $s>0$ (and for arbitrary $a$).  

{\it Proposition 5}.  The function $\gamma_1(a) \to -\infty$ as $a \to 0^+$ and as
$a \to \infty$.  Indeed, the asymptotic form as $a \to \infty$ is $\gamma_1(a) \sim
-{1 \over 2}\ln^2 a$ as can be seen from (2.17).  From Proposition 1(ii) we have
the derivative
$$\gamma_1'(a)=\zeta'(2,a)+\zeta(2,a)=\zeta'(2,a)+\psi'(a)=\sum_{n=0}^\infty {{1-\ln(n+a)} \over {(n+a)^2}},  \eqno(2.33)$$
where the term $\zeta(2,a)=\psi'(a)$ is the trigamma function.  The function
$\gamma_1'(a) \to \infty$ as $a \to 0^+$ and $\to 0$ through negative values
as $a \to \infty$.  The sole zero of $\gamma_1'(a)$ occurs at $a_1^*$, where
$\gamma_1(a_1^*) > 0$.  Therefore, $\gamma_1$ has the single global maximum as
claimed.

{\bf Remark}.  The approximate value $\gamma_1(a_1^*) \simeq 0.0379557$.

{\it Proposition 6}.  As described below, we have for Re $s<1$
$$\int_0^1 \zeta^{(j)}(s,a)da=0.  \eqno(2.34)$$
We differentiate (2.8) $j$ times with respect to $s$ and apply this equation,
putting $z=1-s$, giving the Proposition.

That (2.34) holds can be seen by first noting that by the functional equation of
the Hurwitz zeta function we have for Re $s<1$  and $0<a<1$ 
$$\zeta(s,a)=2^s \pi^{s-1} \Gamma(1-s)\sum_{n=1}^\infty \sin\left(2\pi n a+{{\pi s}
\over 2}\right) n^{s-1}.  \eqno(2.35)$$
By using the product rule, various forms of $\zeta^{(j)}(s,a)$ follow, and these
give (2.34).  

{\bf Remarks}.  Equation (2.34) could also be found on the basis of various 
integral representations for $\zeta$ holding for Re $s<1$.  

Proposition 6 gives a generalization of Proposition 4 of \cite{coffeyjmaa} when
$j>0$.

As a byproduct of our proof of Proposition 6 we have
{\newline \bf Corollary 4}.  We have
$$\sum_{k=0}^\infty {{\gamma_{k+n}(a)} \over {k!}}=(-1)^n [\zeta^{(n)}(0,a)+n!].
\eqno(2.36)$$
This is an extension of the ``beautiful sum" at $a=1$ attributed to O. Marichev \cite{eric}.

As a special case, we have
{\newline \bf Corollary 5}.  We have
$$\sum_{k=0}^\infty {{\gamma_{k+1}(a)} \over {k!}}=-\zeta'(0,a)-1
={1 \over 2}\ln (2\pi) -\ln \Gamma(a)-1.  \eqno(2.37)$$

Indeed, Corollary 4 is consistent with the relation (cf. \cite{coffeyjmaa}, p. 610)
$R_j(a)=-(-1)^j \zeta^{(j)}(0,a)$, where
$${{R'_{j+1}(a)} \over {j+1}}= -\gamma_j-{1 \over a}\ln^j a-\sum_{n=1}^\infty
\left[{{\ln^j(n+a)} \over {n+a}}-{{\ln^j a} \over n}\right], ~~~~ j \geq 0. \eqno(2.38)$$
By integrating this equation we have
$$R_j(a)-R_j(1)=j(1-a)\gamma_{j-1}-\ln^j a-\sum_{n=1}^\infty \left[\ln^j(n+a)-j 
{{\ln^{j-1} n} \over n}(a-1)\right].  \eqno(2.39)$$

We record an expression for $(-1)^{j+1}R_j(a)$ in the following.
{\newline \bf Lemma 2}.  We have for integers $j \geq 1$
$$\zeta^{(j)}(s,a)=(-1)^j a^{1-s}\sum_{k=0}^j {j \choose k}(j-k)! {{\ln^k a} \over {(s-1)^{j-k+1}}}+{{(-1)^j} \over 2}a^{-s} \ln^j a$$
$$+(-1)^j \int_0^\infty {{P_1(x)} \over {(x+a)^{s+1}}}\ln^{j-1} (x+a) ~dx
-(-1)^j s\int_0^\infty {{P_1(x)} \over {(x+a)^{s+1}}}\ln^j (x+a) ~dx,  \eqno(2.40)$$
giving
$$\zeta^{(j)}(0,a)=a\sum_{k=0}^j {j \choose k}(j-k)!(-1)^{k+1}\ln^k a+{{(-1)^j} \over
2} \ln^j a+(-1)^j j\int_a^\infty {{\ln^{j-1} x} \over x}P_1(x-a)dx.  \eqno(2.41)$$
Lemma 2 follows from a direct calculation using the integral representation \cite{yue} (2.3) valid for Re $s >-1$,
$$\zeta(s,a)={a^{-s} \over 2}+{a^{1-s} \over {s-1}}-s\int_0^\infty {{P_1(x)} \over {(x+a)^{s+1}}}dx. \eqno(2.42)$$
Equation (2.40) may also be proved by induction.  The $a=1$ reduction of (2.42) is
well known \cite{titch} (p. 14).

{\it Proposition 7}.  We have from \cite{yue} (pp. 153-154)
$$C_n(a)=\int_1^\infty P_1(x-a){{\ln^{n-1} x} \over x^2}(n-\ln x)dx.  \eqno(2.43)$$
We put $e_n \equiv \exp(n/2)$.  We note that the generic function $\ln^n x/x^2$ is
nonnegative and monotonically increasing for $1<x<e_n$ and nonnegative and
monotonically decreasing for $e_n<x <\infty$.  We have that $P_1(x-a)$ is bounded
and integrable and we split the integrals in (2.43) at $e_{n-1}$ and $e_n$.
By the second mean value theorem for integrals (e.g., \cite{grad}, pp. 1097-98) we
have
$$\int_1^{e_n} {{\ln^n x} \over x^2}P_1(x-a)dx={{\ln^n e_n} \over e_n^2}\int_\eta^{e_n}
P_1(x-a)dx, \eqno(2.44)$$
for some $\eta$ with $1 \leq \eta \leq e_n$.  Therefore, we obtain
$$\int_1^{e_n} {{\ln^n x} \over x^2}P_1(x-a)dx={n^n \over {2^n e^n}}\int_\eta^{e_n}
P_1(x-a)dx \leq {1 \over 6}{n^n \over {2^n e^n}}   
. \eqno(2.45)$$
Here we have used the standard Fourier series for $P_1(x)$ \cite{yue} (p. 151) or
\cite{nbs} (p. 805),
$$P_1(x)=-\sum_{n=1}^\infty {{\sin(2n\pi x)} \over {n\pi}},   \eqno(2.46)$$
giving
$$\int_b^c P_1(x-a)dx={1 \over 2}\left. \sum_{n=1}^\infty {{\cos[2n\pi (x-a)]} \over {n^2\pi^2}}\right|_b^c, \eqno(2.47)$$
so that
$$\left|\int_b^c P_1(x-a)dx\right| \leq {2 \over {2\pi^2}}\zeta(2)={1 \over 6}.  \eqno(2.48)$$
Similarly, we have
$$\int_{e_n}^\infty {{\ln^n x} \over x^2}P_1(x-a)dx={{\ln^n e_n} \over e_n^2}\int_{e_n}^\xi P_1(x-a)dx, \eqno(2.49)$$
for some $\xi$ with $e_n \leq \xi < \infty$.   This gives
$$\int_{e_n}^\infty {{\ln^n x} \over x^2}P_1(x-a)dx={n^n \over {2^n e^n}}\int_{e_n}^\xi P_1(x-a)dx \leq {1 \over 6}{n^n \over {2^n e^n}}       
. \eqno(2.50)$$
Combining the $4$ integral contributions of (2.43) yields the Proposition, as
$$|C_n(a)| \leq {1 \over 3}\left[{{(n-1)^{n-1}n} \over {2^{n-1}e^{n-1}}} + {n^n \over
{2^n e^n}}\right] \leq {{en^n} \over {2^n e^n}}.  \eqno(2.51)$$

{\bf Remarks}.  
Zhang and Williams previously found the better bound
$$|C_n(a)| \leq {{[3+(-1)^n](2n)!} \over {n^{n+1} (2\pi)^n}}.  \eqno(2.52)$$
However, our presentation shows possibility for improvement.  If, for instance,
a better bound is found for the integrals $\int_b^c P_1(x-a)dx$, then we may
expect a tighter estimation.

Let us note also that the use of (2.40) for $\zeta^{(j)}(pk+1,a)$ together with
the expressions of Proposition 2 provides many other opportunities for the
estimation of $|\gamma_k(a)|$ and $|C_k(a)|$.

{\it Proposition 8}.  We apply the formula of Klusch \cite{klusch} (2.5),
$$L(x,s,a+\xi)=\sum_{k=0}^\infty (-1)^k {{(s)_k} \over {k!}} L(x,s+k,a) \xi^k.
\eqno(2.53)$$ 
This formula may be obtained by Taylor expansion, expansion of an integral
representation, or by binomial expansion in (1.15).  We take $n$ derivatives
of (2.53) using the product rule.  We evaluate at $s=1$ using Lemma 1 and the
defining relation (1.17) of $\ell_n$.  Separating the $k=0$ term yields the
Proposition and then Corollary 3.

{\bf Remark}.  As very special cases, we have
$$\ell_0\left({1 \over 2},a\right)={1 \over 2}\left[\psi\left({{1+a} \over 2}\right)-
\psi\left({{a} \over 2}\right)\right],$$
$$\ell_1\left({1 \over 2},a\right)={1 \over 2}\left\{\ln2 \left[\psi\left({{a} \over 2}\right)- \psi\left({{1+a} \over 2}\right)\right]+\gamma_1\left({{1+a} \over 2}\right)-\gamma_1\left({{a} \over 2}\right) \right\},$$
and
$$\ell_2\left({1 \over 2},a\right)={1 \over 4}\left\{\ln^2 2 \left[\psi\left({{1+a} \over 2}\right)- \psi\left({{a} \over 2}\right)\right]+2\ln2 \left[\gamma_1\left({{a} \over 2}\right)-\gamma_1\left({{a+1} \over 2}\right) \right] \right.$$
$$\left. + \gamma_2\left({{a} \over 2}\right)-\gamma_2\left({{a+1} \over 2}\right)  \right\}.  \eqno(2.54)$$

As another example, from
$$L\left({1 \over 4},s,a\right)=\sum_{n=0}^\infty {i^n \over {(n+a)^s}}=4^{-s}\left
\{\zeta\left(s,{a \over 4}\right)-\zeta\left(s,{{a+2} \over 4}\right)+i\left[
\zeta\left(s,{{a+1} \over 4}\right)-\zeta\left(s,{{a+3} \over 4}\right)\right]
\right\}, \eqno(2.55)$$
by taking the real and imaginary parts of derivatives evaluated at $s=1$, we 
obtain expressions for $\gamma_k(a/4)-\gamma_k[(a+2)/4]$ and $\gamma_k[(a+1)/4]-\gamma_k[(a+3)/4]$.  Evidently for $x$ a rational number such 
relations always exist.  

{\it Proposition 9}.  From the classical theta function-based representation of the
Riemann zeta function \cite{coffey04} (2) we have
$$\Gamma(s/2)\zeta(s)={\pi^{s/2} \over {s(s-1)}}+\sum_{n=1}^\infty n^{-s}\Gamma\left(
{s \over 2},\pi n^2\right) + \pi^{s-1/2}\sum_{n=1}^\infty n^{s-1}\Gamma\left(
{{1-s} \over 2},\pi n^2\right).  \eqno(2.56)$$
We expand both sides of this equation in powers of $s-1$, and equate the 
coefficients of $(s-1)^0$ and $(s-1)^1$ on both sides to obtain parts (i) and (ii),
respectively.  The coefficient of $(s-1)^{n-1}$ for $n \geq 0$ of the term
$\pi^{s/2}/[s(s-1)]$ is given by $\pi^{1/2}(-1)^n\sum_{j=0}^n (-1)^j\ln^j \pi/2^j$,
and the simple polar term from the left side of (2.56) is cancelled by the $n=0$
term.  We use the special function relations $\Gamma(0,z)=-\mbox{Ei}(-z)$ and
$\Gamma(1/2,z)=\sqrt{\pi}[1-\mbox{erf}(\sqrt{z})]$.  
The incomplete Gamma function \cite{grad} is given for Re $x>0$ by$$\Gamma(\alpha,x)=\int_x^\infty e^{-t}t^{\alpha-1}dt={{2x^\alpha e^{-x}}\over {\Gamma(1-\alpha)}}\int_0^\infty{{t^{1-2\alpha}e^{-t^2}} \over {t^2+x}}dt, \eqno(2.57)$$
where the latter form holds for Re $\alpha < 1$.  It is convenient in finding 
derivatives of this function to use the relation$$\Gamma(\alpha,x) = \Gamma(\alpha)-{x^\alpha \over \alpha} ~_1F_1(\alpha;\alpha+1;-x), ~~~~~~ -\alpha \notin N, \eqno(2.58)$$
where $_1F_1$ is the confluent hypergeometric function \cite{nbs,grad}.  
It follows that 
$${d \over {d\alpha}} ~_1F_1(\alpha;\alpha+1;-x)={1 \over \alpha}\left [_1F_1(\alpha;\alpha+1;-x) - ~_2F_2(\alpha,\alpha;\alpha+1,\alpha+1;-x) \right ],\eqno(2.59)$$
and
$${d \over {d\alpha}}\Gamma(\alpha,x)=\Gamma(\alpha)\psi(\alpha)+{x^\alpha\over \alpha^2}\left [-\alpha \ln x ~_1F_1(\alpha;\alpha+1;-x) + ~_2F_2(\alpha, \alpha;\alpha+1,\alpha+1;-x) \right ]. \eqno(2.60)$$   
When $\alpha \to 0$ in (2.60), the singular terms $1/\alpha^2$ cancel.  In fact,
we have the expansions
$$\Gamma'(\alpha)=\Gamma(\alpha)\psi(\alpha)=-{1 \over \alpha^2}+{\gamma^2 \over 2}
+{\pi^2 \over {12}}+O(\alpha), \eqno(2.61)$$
and
$${x^\alpha \over \alpha^2}{{(\alpha)_j} \over {(\alpha+1)_j}}\left[-\alpha \ln x
+{{(\alpha)_j} \over {(\alpha+1)_j}}\right] {{(-x)^j} \over {j!}}={{(1-j \ln x)}
\over {j^3 (j-1)!}}(-x)^j + O(\alpha), ~~~~ \alpha \to 0.  \eqno(2.62)$$
Summing the latter relation on $j$ as $\alpha \to 0$ then gives
$$\sum_{j=1}^\infty {{(1-j \ln x)} \over {j^3 (j-1)!}}(-x)^j =-x ~_3F_3(1,1,1;2,2,2;-x)
+[\gamma+\Gamma(0,x)+\ln x]\ln x.  \eqno(2.63)$$
We apply these relations at $\alpha=0$ and $1/2$ and the Proposition follows.

{\bf Remarks}.  By making a change of variable in the theta function-based representation \cite{coffey04} (2) we may include a parameter in the series representation (2.56), and so in Proposition 9 too.  Beyond this, we may include
a parameter $b>0$ and a set of polynomials $p_j(s)$ with zeros lying only on the
critical line in representing the Riemann zeta and xi functions \cite{coffeypla}
(Proposition 3).  Thereby, we obtain a generalization of Proposition 9.

In equation (1.20), the sum terms provide a small correction $\simeq 0.0230957$ to produce the value $\gamma/2 \simeq 0.288607$.  Although there are infinite sum
corrections in Proposition 9, the expressions such as (1.20) and (1.21) may have
some attraction for computation.  This is owing to the very fast decrease of the
summands with $n$.  For (1.20), using known asymptotic forms, the summand terms
have exponential decrease $\sim e^{-n^2\pi}\left[{2 \over {\pi n^2}}+O\left({1 \over 
n^4}\right)\right]$.  Similarly for (1.21), the summand has exponential decrease 
in $n$.  Therefore, high order approximations for
the constants may be obtained with relatively few terms.

We note an integral representation for the term $-\sum_{n=1}^\infty \mbox{Ei}(-\pi
n^2)$ in Proposition 9 in the following.
{\newline \bf Lemma 3}.  Put the function $\theta_3(y)=1+2\sum_{n=1}^\infty y^{n^2}$.
Then we have
$$-\sum_{n=1}^\infty \mbox{Ei}(-\pi n^2)=-{1 \over 2}\int_0^1 \left[\theta_3\left(
e^{\pi/(x-1)}\right)-1\right]{{dx} \over {x-1}}$$
$$=-{1 \over 2}\int_1^\infty [1-\theta_3(e^{-\pi u})]{{du} \over u}.  \eqno(2.64)$$

Proof.  Let $L_n^\alpha$ be the Laguerre polynomial of degree $n$ and parameter
$\alpha$.  We use the relation (e.g., \cite{grad}, p. 1038)
$$\Gamma(\alpha,x)=x^\alpha e^{-x} \sum_{k=0}^\infty {{L_k^\alpha(x)} \over {k+1}},
~~~~~~ \alpha > -1, \eqno(2.65)$$
at $\alpha=0$.  We then have
$$-\sum_{n=1}^\infty \mbox{Ei}(-\pi n^2)=\sum_{n=1}^\infty e^{-\pi n^2} \sum_{k=0}^
\infty {{L_k(\pi n^2)} \over {k+1}}$$
$$=\sum_{n=1}^\infty e^{-\pi n^2} \sum_{k=0}^\infty \int_0^1 x^k dx \sum_{k=0}^
\infty L_k(\pi n^2)$$
$$=-\sum_{n=1}^\infty e^{-\pi n^2}\int_0^1 {e^{x \pi n^2/(x-1)} \over {x-1}}dx,
\eqno(2.66)$$
where we used the generating function of the Laguerre polynomials (e.g., \cite{grad}, p. 1038).  The interchange of summation and integration is justified by the absolute
convergence of the integral.  Using the definition of $\theta_3$ completes the Lemma.

\medskip
\centerline{\bf Discussion:  Generalization of Proposition 9}
\medskip

We may generalize Proposition 9 to the context of generalized Stieltjes constants
coming from the evaluation of Dirichlet $L$ series derivatives at $s=1$.  For this
we require a number of definitions.  We use \cite{karatsuba} (Ch. 1, Section 4.2).

Let $\chi$ be a primitive character modulo $k$.  We need two theta functions,
$$\theta(\tau,\chi)=\sum_{n=-\infty}^\infty \chi(n) e^{-\pi \tau n^2/k}, $$
for $\chi$ an even character, 
$$\theta_1(\tau,\chi)=\sum_{n=-\infty}^\infty n\chi(n) e^{-\pi \tau n^2/k}, $$
for $\chi$ an odd character, and the Gauss sum
$$g(\chi)=\sum_{j=1}^k \chi(j)e^{2\pi ij/k}.$$
We let $\delta= 0$ or $1$ depending upon whether $\chi(-1)=1$ or $-1$, respectively.
We put
$$\xi(s,\chi)=(\pi k^{-1})^{-(s+1)/2}\Gamma\left({{s+\delta} \over 2}\right)
L(s,\chi).$$

Then by \cite{karatsuba} (p. 15) we find for $\chi$ an even character,
$$2\xi(s,\chi)=\int_1^\infty \tau^{s/2-1}\theta(s,\chi)d\tau+{\sqrt{k} \over {g(\bar{\chi})
}}\int_1^\infty \tau^{-(s+1)/2}\theta(\tau,\bar{\chi})d\tau$$
$$=\pi^{-s/2-1}k^{s/2+1}\sum_{n=-\infty}^\infty \chi(n)n^{-s-2} \Gamma\left({s \over 2}
+1,\pi {n^2 \over k}\right)$$
$$+{\sqrt{k} \over {g(\bar{\chi})}}\pi^{(s-1)/2}k^{(1-s)/2}
\sum_{n=-\infty}^\infty \bar{\chi}(n)n^{s-1} \Gamma\left({{1-s} \over 2}, \pi {n^2 \over k}\right),$$
and for $\chi$ an odd character,
$$2\xi(s,\chi)=\int_1^\infty \tau^{(s-1)/2}\theta_1(s,\chi)d\tau+{{i\sqrt{k}} \over {g(\bar{\chi})}}\int_1^\infty \tau^{-s/2}\theta_1(\tau,\bar{\chi})d\tau$$
$$=\pi^{-(s+1)/2}k^{(s+1)/2}\sum_{n=-\infty}^\infty n\chi(n)n^{-(s+1)} \Gamma\left({{s +1} \over 2}+1,\pi {n^2 \over k}\right)$$
$$+{{i\sqrt{k}} \over {g(\bar{\chi})}}\pi^{s/2-1}k^{1-s/2}\sum_{n=-\infty}^\infty n\bar{\chi}(n)n^{s-2} \Gamma\left(1-{s \over 2}, \pi {n^2 \over k}\right).$$
Then one may multiply the expressions for $\xi(s,\chi)$ by $(\pi k^{-1})^{(s+1)/2}$
and proceed as in the proof of Proposition 9 in developing both sides in powers of
$s-1$.

In the case of the Hurwitz zeta function we employ the theta function
$$\theta(\tau,a)=\sum_{n \neq 0} e^{-\pi \tau(n+a)^2},$$
with its functional equation
$$\theta(1/\tau,a)=\sqrt{\tau}e^{-\pi a^2/\tau}\theta(\tau,-ia/\tau).$$

{\it Proposition 10}.  We make use of expressions in the fine paper of Fine 
\cite{fine}.  In particular, see his expressions for the function $H(a,s)$
(pp. 362-363).  We have
$$\pi^{(s-1)/2}\Gamma\left({{1-s} \over 2}\right)[\zeta(1-s,a)+\zeta(1-s,1-a)]
=2\sum_{n=1}^\infty \cos(2\pi n a)\int_0^\infty e^{-\pi n^2 t}t^{s/2-1}dt$$
$$=2\pi^{-s/2}\Gamma\left({{s} \over 2}\right) \sum_{n=1}^\infty {{\cos(2\pi n a)}
\over n^s}$$
$$=\pi^{-s/2}\Gamma\left({{s} \over 2}\right)[\mbox{Li}_s(e^{2\pi ia})+
\mbox{Li}_s(e^{-2\pi ia})].  \eqno(2.67)$$
Initially, the left side of this equation is valid for Re $s<0$.  With analytic
continuation, the function is $H(a,s)+2/s$ is entire in $s$.  Therefore, we may
expand both sides of the representation (2.64) about $s=0$.  As it must, the singular
term $-2/s$ effectively cancels from both sides of (2.67).  Part (i) of the
Proposition results from the constant term $s^0$ on both sides of (2.67), and
using the reflection formula for the digamma function $\psi(1-a)=\psi(a)+\pi \cot \pi a$.  Similarly, part (ii) results from the term $s^1$ on both sides of (2.60).

{\bf Remarks}.  Proposition 10 and (2.67) properly reduce as they should for
$a=1/2$.  In this case, we have the alternating zeta function $\sum_{n=1}^\infty
{{(-1)^n} \over n^s}=(2^{1-s}-1)\zeta(s)$.  Then both sides of part (i) of
Proposition 10 yield $\gamma-\ln \pi +2 \ln 2$.

{\it Proposition 11}.  By using the definition (1.24) we have
$$P_{n+1}(y)={1 \over {n!}}\int_0^y x(1-x)_{n-1} ~dx$$
$$={1 \over {n!}}\sum_{k=0}^{n-1} (-1)^{n+k-1} s(n-1,k) \int_0^y x(1-x)^k dx$$
$$={{(-1)^n} \over {n!}}\sum_{k=0}^{n-1} (-1)^{k-1} {{s(n-1,k)} \over {(k+2)(k+1)}}
[(ky+y+1)(1-y)^k(y-1) + 1], \eqno(2.68)$$
whence part (i) follows.  Of the many ways to perform the elementary integral here,
it may be evaluated as a special case of the incomplete Beta function (e.g., 
\cite{grad}, p. 950)
$$B_x(p,q)=\int_0^x t^{p-1}(1-t)^{q-1}dt={x^p \over p} ~_2F_1(p,1-q;p+1;x), ~~~~~
\mbox{Re} ~p >0, ~~~\mbox{Re} ~q >0.  \eqno(2.69)$$ 

For part (ii) we again use a generating function relation for $s(n,k)$, writing
$$P_{n+1}(y)=-{1 \over {n!}}\int_0^y (-x)_n ~dx={{(-1)^{n+1}} \over {n!}}\sum_{k=0}^n
s(n,k)\int_0^y x^k dx.  \eqno(2.70)$$
The second line of (1.26) simply follows from the values $s(n,0)=\delta_{n0}$ and
$s(n,n)=1$.

The identity of part (iii) follows simply from putting $y=1$ in parts (i) and (ii).

For part (iv), we have the well known expression
$$\gamma_k = \lim_{N \to \infty}\left(\sum_{m=1}^N {{\ln^k m} \over m}-{1 \over {k+1}}\ln^{k+1} N\right), \eqno(2.71)$$
so that $\lim_{n \to \infty} r_n^{(k)} = 0$.  We form
$$D_n^{(k)}=\sum_{m=1}^n \left({{\ln^k m} \over m}-\int_m^{m+1} {{\ln^k x} \over x}
dx\right)$$
$$=\sum_{m=1}^n \int_0^1 \left[{{\ln^k m} \over m}-{{\ln^k (x+m)} \over {x+m}}\right]
dx.  \eqno(2.72)$$
Therefore, we have
$$r_n^{(k)}=\sum_{m=n+1}^\infty \int_0^1 \left[{{\ln^k m} \over m}-{{\ln^k (x+m)} 
\over {x+m}}\right] dx.  \eqno(2.73)$$
We now use \cite{coffeystdiffs} (5.7) 
$${{\ln^n y} \over y}-{{\ln^n (x+y)} \over {x+y}}=\sum_{k=1}^\infty {x^k \over y^{k+1}}\left[(-1)^k\ln^n y + {1 \over {k!}}\sum_{j=0}^{n-1} {{n!} \over {(n-j-1)!}}
s(k+1,j+2) \ln^{n-j-1} y \right].  \eqno(2.74)$$
Performing the integration of (2.73) gives part (iv).

\medskip
\centerline{\bf Discussion related to Proposition 11}
\medskip

By defining the function $f_k(x,m)=(x+m)\ln^k m-m\ln^k(x+m)$, it is possible to
further decompose the remainders $r_n^{(k)}$ of (1.29) by writing
$$r_n^{(k)}=\sum_{m=n+1}^\infty \int_0^1 f_k(x,m)\left[{1 \over {m(x+m)}}-{1 \over
{m(m+1)}}\right]dx+\int_0^1 \sum_{m=n+1}^\infty {{f_k(x,m)} \over {m(m+1)}}dx.
\eqno(2.75)$$
The representation (2.74) can then be used three times in this equation.  This
process of adding and subtracting terms can be continued, building an integral
term with denominator $m(m+1)\cdots(m+j)(m+x)$.  For $k=0$, there is drastic
reduction to the original construction of Ser \cite{ser}.

Proposition 11 (iv) can be easily extended to expressions for $r_n^{(k)}(a) =\gamma_k(a)-D_n^{(k)}(a)$.

As pointed out by Ser, it is possible to write many summations and generating 
function relations with the polynomials $P_n$.  As simple examples, we have
$$\sum_{n=1}^\infty P'_{n+1}(y)z^n=-\sum_{n=1}^\infty (-1)^n{y \choose n}z^n
=1-(1-z)^y, \eqno(2.76)$$
and
$$\sum_{n=1}^\infty {{P'_{n+1}(y)} \over n}=-\sum_{n=1}^\infty {{(-1)^n} \over n}{y \choose n}=\psi(y+1)+\gamma=H_y. \eqno(2.77)$$
With the form (1.27) we easily verify a generating function:
$$\sum_{n=1}^\infty p_{n+1}z^{n-1}=\sum_{n=1}^\infty {{(-1)^{n+1}} \over {n!}} \sum_{k=1}^n {{s(n,k)} \over {k+1}}z^{n-1}$$
$$=-{1 \over z}\sum_{k=1}^\infty {1 \over {k+1}}\sum_{n=k}^\infty {{s(n,k)} \over
{n!}}(-z)^n=-{1 \over z}\sum_{k=1}^\infty {1 \over {(k+1)!}}\ln^k(1-z)$$
$$={1 \over z}+{1 \over {\ln(1-z)}}.  \eqno(2.78)$$
One may also write a great many hypergeometric series of the form $\sum_{n=1}^\infty
{{P_{n+1}'(y)} \over n^j}z^n$, that we omit.

In contrast to an expression of Ser \cite{ser}, we have
$$\sum_{n=2}^\infty p_{n+1}(1-e^z)^{n-1}={1 \over z}+{1 \over {1-e^z}}-{1 \over 2}$$
$$={1 \over z}-{1 \over 2} \coth\left({z \over 2}\right)$$
$$=\sum_{k=1}^\infty {{B_{2k}} \over {(2k)!}} z^{2k-1},  ~~~~|z| < 2\pi, \eqno(2.79)$$
where the latter series may be found from \cite{grad} (p. 35).  This relation
serves to connect the numbers $p_{n+1}$ with the values $\zeta(2k)$.
Equivalently, using an integral representation for $\zeta(2k)$, we have
$$\sum_{n=2}^\infty p_{n+1}(1-e^z)^{n-1}=2\int_0^\infty \sinh(zv){{dv} \over
{e^{2\pi v}-1}}.  \eqno(2.80)$$
As well, we have many extensions including
$$\sum_{n=2}^\infty p_{n+1}(1-e^z)^{n-1}e^{-tz(n-1)}={1 \over {\ln[1+e^{z(1-t)}-e^{-tz}]}}+{e^{tz} \over {1-e^z}}-{1 \over 2}.  \eqno(2.81)$$

We may inquire as to the asymptotic form of $p_{n+1}$ as $n \to \infty$.  Using
only the leading term of the result of \cite{wilf}, we have
$$p_{n+1} \sim -{1 \over n}\sum_{k=1}^n{{(-1)^k} \over {(k+1)}}{{\ln^{k-1} n} \over
{(k-1)!}} \sim {1 \over {n \ln^2 n}}, ~~~~ n \to \infty.  \eqno(2.82)$$
There are corrections to this result which overall decrease it.  Therefore, one
may wonder if it can be proved that the expression on the right serves as an upper
bound for all $n \geq 2$.  

Motivated by \cite{nbs} (p. 824), we conjectured that
$$p_{n+1} \sim {1 \over {n (\ln n+\gamma)^2}}, ~~~~ n \to \infty,  \eqno(2.83)$$
is an improved asymptotic form and upper bound.  This asymptotic form has been
verified by Knessl \cite{knessl}.  In fact, he has obtained a full asymptotic
series for $p_{n+1}$ in the form
$$p_{n+1} \sim {1 \over {n \ln^2 n}}\left(1+\sum_{j=1}^\infty {A_j \over 
{\ln^j n}}\right), \eqno(2.84)$$
where $A_j$ are constants that may be explicitly determined from certain
logarithmic-exponential integrals.  We have $A_1=-2\gamma$ and $A_2=3\gamma^2
-\pi^2/2$.  Knessl develops (2.84) from the exact representation
$$p_{n+1}=\int_0^\infty {1 \over {(1+u)^n}} {{du} \over {(\ln^2 u+\pi^2)}},
~~~~~~ n \geq 1.  \eqno(2.85)$$

From (2.85) we have the following
\newline{\bf Corollary 6}.  We have the integral representation
$$\gamma=\int_{-\infty}^\infty e^z {{\ln(1+e^{-z})} \over {z^2 +\pi^2}}dz.  \eqno(2.86)$$
{\em Proof}.  We have \cite{grad} (p. 943)
$$\gamma=-\psi(1)=\int_0^1 \left({1 \over {\ln u}}+{1 \over {1-u}}\right)du
=\int_0^1 \left({1 \over {\ln (1-v)}}+{1 \over v}\right)dv$$
$$=\int_0^1 \sum_{n=1}^\infty p_{n+1} v^{n-1}dv=\sum_{n=1}^\infty {p_{n+1}
\over n}, \eqno(2.87)$$
where we used (2.78).  Now we use Knessl's representation (2.85), giving
$$\gamma=\sum_{n=1}^\infty {1 \over n}\int_0^\infty {1 \over {(1+u)^n}} {{du} 
\over {(\ln^2 u+\pi^2)}}$$
$$=-\int_0^\infty \ln\left({u \over {1+u}}\right)  {{du} \over {(\ln^2 u+\pi^2)}}$$
$$=-\int_{-\infty}^\infty  {e^z \over {(z^2+\pi^2)}}[z-\ln(1+e^z)]dz$$
$$=\int_{-\infty}^\infty  {e^z \over {(z^2+\pi^2)}}\ln(1+e^{-z})dz.  \eqno(2.87)$$

\medskip
\centerline{\bf Summary and very brief discussion}
\medskip

Our methods apply to a wide range of functions of interest to special function
theory and analytic number theory, including but not limited to, the Lerch zeta
function and Dirichlet $L$ functions.  Two dimensional extensions would be to Epstein and double zeta functions.  Our results include an addition formula for the Stieltjes coefficients, as well as expressions for their derivatives.  The series representations for $\gamma_k(a)$ have very rapidly convergent forms, making them applicable for multiprecision computation.  Byproducts of our results include some summation
relations for the Stieltjes coefficients.  We have given a means for estimating
the magnitude of these coefficients, although this remains an outstanding problem.

\medskip
\centerline{\bf Acknowledgement}
\medskip

I thank R. Smith for useful correspondence.  I thank C. Knessl for reading
the manuscript and for access to his results on the constants $p_{n+1}$.

\pagebreak

\end{document}